# Analogy Electromagnetism - Acoustics: Validation and Application to Local Impedance Active Control for Sound Absorption

L. Nicolas
CEGELY - UPRESA CNRS 5005 - Ecole Centrale de Lyon
BP163 - 69131 Ecully cedex - France

M. Furstoss, M.A. Galland
LMFA - UMR CNRS 5509 - Ecole Centrale de Lyon
BP163 - 69131 Ecully cedex - France

*Abstract*– An analogy between electromagnetism and acoustics is presented in 2D. The propagation of sound in presence of absorbing material is modeled using an open boundary microwave package. Validation is performed through analytical and experimental results. Application to impedance active control for sound absorption is finally described.

I. INTRODUCTION

The original form of active noise control consisted in sound pressure reduction by the superposition of an acoustic signal 180 degrees out of phase with the unwanted sound [1]. The additional sound is emitted by a secondary source. This method provides an attractive low frequency alternative to the inefficient passive treatments of using absorbent materials. In recent years, due to advances in digital signal processing hardware, active control technology has become an industrial reality.

By locating the active "noise reducer" behind a resistant cloth, the acoustic dissipation is increased by friction in the low frequency range [2]. This method appears to be efficient in two main cases. The first one concerns the reduction of the low frequency noise radiated by inlet or exhaust ducts. In this case, it is almost impossible to apply active control methods based on interference principle because no perfect correlation exists between primary and secondary waves. For such a case, active absorption achieved along the duct walls by local control set-ups is a very promising alternative. The second case concerns the active sound absorption for noise reduction in enclosed sound fields. A general formulation has been proposed for this problem by considering two different approaches [3]: local and global control of sound. In the first of these strategies, the aim remains to achieve pressure releases at a set of discrete locations. No attention is paid to the overall acoustic field in the volume. Applications of this strategy concern for example the noise reduction at the passenger heads in car or propeller aircraft cabins. The second approach, on the contrary, attempts to minimize the overall acoustic level ( the total acoustic energy inside the cavity). Nevertheless, experimental difficulties (especially concerning the energy estimation) limit its application to realistic environments. A good alternative consists in changing the boundary conditions and thus the wall impedances to reduce the total acoustic energy.

Various set-ups have been developed by many research teams to achieve active control of the impedance [4], which is defined on a surface by the ratio between pressure and normal velocity. Two different strategies have been proposed: the first one is named direct impedance control because it involves sensors to estimate both pressure and velocity close to the secondary source. However, its implementation is limited by major drawbacks such as bandwidth limitations and difficulties for extending absorbing surfaces. The second strategy combines active control and porous material, in order to improve its absorbent behavior. In a previous study [4] we have highlighted that this association allows the active control to be reduced to a simple pressure release if the porous layer characteristics are well suited. These hybrid passive/active systems then enable feedback control techniques to be implemented and extended absorbing coatings to be achieved.

In order to optimize such absorbers and their location, attention is now paid to precisely modelize their behavior. Only numerical computation can simulate the acoustic propagation in unbounded domains in the presence of a porous material. Boundary elements methods have been extensively used to determine pressure fields in radiation problems. But up to now, absorption is generally taken into account by introducing impedance boundary conditions. This cannot exactly modelize propagation phenomena occurring in porous media. On the other hand, modeling sound absorbing material has been developed with a good accuracy only over the past ten years. The adaptation of a 2D package previously developed for open boundary microwaves [5] has allowed us to associate porous material and free field radiation.

The objective of this paper is to show how an analogy between acoustics and electromagnetism has made possible the use of this package to modelize our acoustic problem. In a first section, we present the analogy used between acoustics and electromagnetism. The second section deals with the validation of the analogy by first studying the simple case of a plane wave impinging on a porous layer in a wave guide. A numerical simulation of the set-up used to measure the surface impedance of a porous layer in free field is also presented to validate the analogy for unbounded domain. The last section deals with the development of large surfaces of active absorbent and show the interest of using numerical simulations to optimize this set-up.

## II. 2D Analogy Acoustics - Electromagnetism

The acoustic behavior of a material layer has to be taken into account with the use of propagation models in porous media. Many classical materials are assumed to have a motionless frame. Thus, they are replaced on a macroscopic scale by an equivalent fluid having a complex dynamic density $\rho_e$ and a complex dynamic bulk modulus K. Solving Helmholtz equation enables their acoustic behavior to be described by a complex wave number k and a complex characteristic impedance Zc. The frequency dependence of these parameters is given by modeling. Usually the model proposed by Allard [6] is used because it is well suited for low frequencies. In this model $\rho_e$ and K are functions of five physical characteristics of the material. For very classical porous absorbers such as fiberglass, this number may be reduced to one single, the resistivity.

In the harmonic case, the acoustic pressure verifies the Helmholtz's equation (1), which can be compared to the scalar wave equation for the magnetic field $H_z$ in TM polarization:

$$\Delta p + \frac{\omega^2 \rho_e(\omega)}{K(\omega)} p = 0 \equiv \Delta H_z + \omega^2 \mu\varepsilon H_z = 0 \quad (1)$$

This leads to the analogy presented in Table I, where the acoustic pressure p is equivalent to the magnetic field $H_z$. The acoustic velocity **v** is then in quadrature with the electric field E. On the contrary of the usual analogy (p≡$E_z$ [7]), this allows the boundary conditions to be equivalent for a perfectly reflecting boundary in acoustics (**n.v**=0) and for a perfect electric conductor (**nxE**=0). Equivalence can also naturally be found between the acoustic intensity vector and the Poynting vector, between the acoustic and the electromagnetic energies, and between the surface impedances.

There is no reason for this analogy to be not valid in 3D. However, the different nature of the fields -scalar in acoustics and vector in electromagnetism- makes the use of an electromagnetic package more complicated in this case.

## III. Validation of the Analogy.

We have developed previously a general 2D Finite Element (F.E.) package for the modeling of open boundary microwave problems [5]. The formulation is given by (2). It handles conducting, permeable, dielectric and absorbing materials. The open boundary is taken into account either by coupling with the boundary element method or by using absorbing boundary condition (Engquist-Majda for a rectangular outer boundary or Bayliss-Turkel for a circular boundary). Second order triangles are used for the F.E. discretization. Far field is computed by boundary element method or by harmonic series expansion [9]. Applications of such a formulation are the design of antennas or the modeling of electromagnetic scattering.

$$\iint_\Omega \left[ \nabla W . \frac{1}{j\omega\varepsilon} \nabla H_z + j\omega\mu W . H_z \right] d\Omega - \frac{1}{j\omega\varepsilon} \int_\Gamma W \frac{\partial H_z}{\partial n} d\Gamma = \iint_\Omega W . J \, d\Omega \quad (2)$$

The same formulation is used for acoustic problems. The analogy is validated using two methods: by comparison with analytical results in the case of a standing wave tube, and by comparison with experimental measurements of the surface normal acoustic impedance of porous materials.

### A. The Standing Wave Tube

The standing wave tube has been used for a long time to measure the reflection coefficient of materials. It is a closed boundary problem. The absorbing material is set at one extremity of the tube, and a plane acoustic wave propagates parallel to the axis of the tube (fig. 1). An analytical solution for this standing wave problem is given in [8]. An excellent agreement is found between analytical and numerical results, either for the pressure (fig. 2) or for the velocity.

TABLE I
ANALOGY BETWEEN ACOUSTIC AND ELECTROMAGNETIC VARIABLES AND MATERIAL CHARACTERISTICS

| acoustic variable | | electromagnetic variable | | analogy |
|---|---|---|---|---|
| P | acoustic pressure | $H_z$ | magnetic field | $p \equiv H_z$ |
| v | acoustic velocity | E | electric field | $v_x \equiv E_y$ and $v_y \equiv -E_x$ |
| $\rho_e(\omega)$ | dynamic density | $\varepsilon' = \varepsilon - j\sigma/\omega$ | complex permittivity | $\rho_e \equiv \varepsilon$ |
| $K(\omega)$ | dynamic bulk modulus | $\mu$ | permeability | $K \equiv 1/\mu$ |

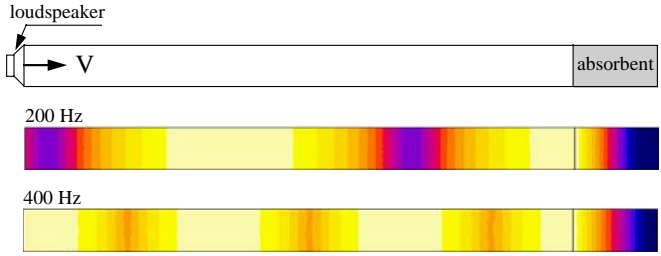

Fig. 1. Geometry of the standing wave tube and acoustic pressure at 200 Hz and 400 Hz.

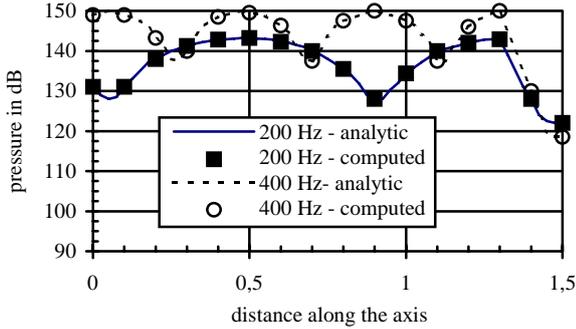

Fig. 2. Analytical and computed pressure along the axis of the tube.

*B. Numerical Simulation of an Impedance Measurement Experimental Set-up*

A plane piston, modelized as a constant speed acoustic source, is located above a fiberglass panel supported by a rigid floor. Fig. 3 shows the computed acoustic pressure when the frequency of the source is 200 Hz and 800 Hz. Obviously, the plane wave hypothesis is more satisfied when the wavelength decreases. The computed surface impedances obtained at the middle of the panel surface are compared in fig. 4. A very good agreement is found from 200 Hz to 1000 Hz with the results provided by the model under plane wave assumption. Moreover, the same impedance value has been computed on the whole surface. Some differences only appear for points located near edges. This is probably due to the proximity of the computation domain boundaries.

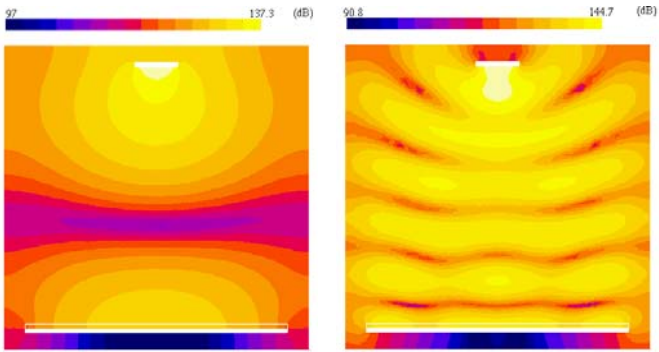

Fig. 3. Acoustic pressure in presence of a porous material. Left: f=200 Hz, right: f=800 Hz

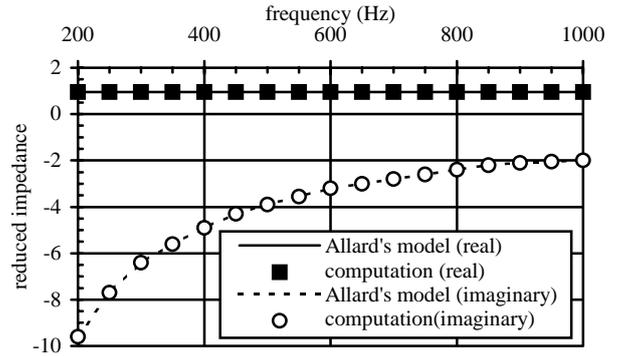

Fig. 4. Normal surface impedance of a porous material.

V. APPLICATION TO THE LOCAL IMPEDANCE ACTIVE CONTROL

*A. Numerical Modeling of the Surface Impedance control of a Porous Material by Pressure Minimization*

A primary source is located above a porous material supported by a rigid floor (fig. 5). A secondary source is located inside the cavity, allowing to cancel the acoustic pressure at the point O. The primary source has a constant velocity $V_0$ and the secondary source has a constant velocity $\lambda V_0$. The global acoustic pressure in any point of the space is then obtained by superposition of the pressures due to both sources ($p_1$: primary, $p_2$: secondary):

$$p(x,y) = p_1(x,y) + \lambda p_2(x,y) \qquad (3)$$

The coefficient $\lambda$ is found by applying the zero pressure condition at the point O. To modelize such a device, three computations are thus necessary: the pressure field separately produced by each source is first computed. This allows to calculate $\lambda$. A third computation cancels the pressure at the point O corresponding to the control microphone position.

A local absorption coefficient may be calculated from the knowledge of the normal impedance on the surface of the absorbent.. Fig. 6 shows this coefficient along the porous surface. A good agreement between the numerical results and the experimental measurements may be found above the secondary source. The absorption is perfect in this area. On the other hand, it seems to be underestimated by the F.E. outside this area. This is certainly due to the 3D nature of the device, and to the anisotropy of the absorbent, which is important for this configuration.

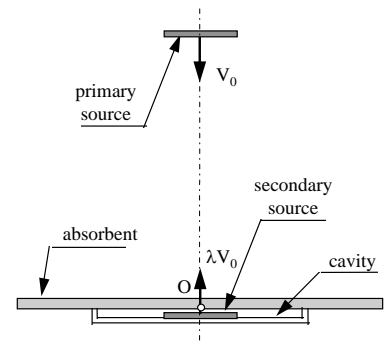

Fig. 5. Experimental set-up for the minimization of the acoustic pressure behind a porous material.

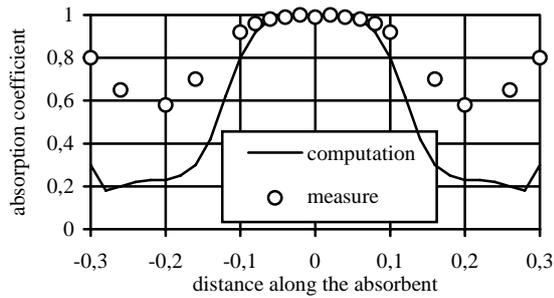

Fig. 6. Absorption coefficients along the absorbent - f = 400 Hz.

*B. Increase of the Active Surface*

Extended absorbing surfaces may be built by assembling elementary active cells behind a large porous material sample. Fig. 7 shows the example of a 3×3 cells array. Each of them includes a loudspeaker which is controlled in order to cancel the pressure at a microphone located at the rear face of the porous material. The 2D FE modeling is performed by considering a single cross section. As previously, the pressure field separately produced by each source is first computed. A linear combination allows then to cancel the pressure at the points corresponding to the control microphone positions. From fig. 8, the influence of active absorption on pressure fields appears clearly: stationary waves produced by the rigid floor are canceled with active control, and the primary source radiates almost as in free field. Experimental results are confirmed by the numerical results (fig. 9). The minimum and maximum values of absorption are especially well predicted by the numerical modeling. Slight differences are due to bad experimental conditions: for example, the absorbent may close incompletely the cells because of the presence of wires.

CONCLUSION

We have shown in this paper how an analogy between two different disciplines of the physic can be made, validated, and used to adapt a numerical formulation. In our case, electromagnetics has been used to design local impedance active sound control systems. Furthermore, we could imagine that, if active control is used for sound absorption, equivalent system could also be used for an electromagnetic compatibility purpose, by reducing electromagnetic radiations with appropriated antennas and sources. Such approach is already used in low frequency (boat demagnetization) and it requires certainly to be studied for higher frequencies.

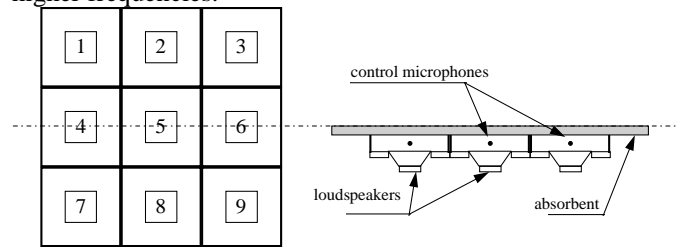

Fig. 7. Control system with nine elementary active cells. Left: front elevation, right: sectional view

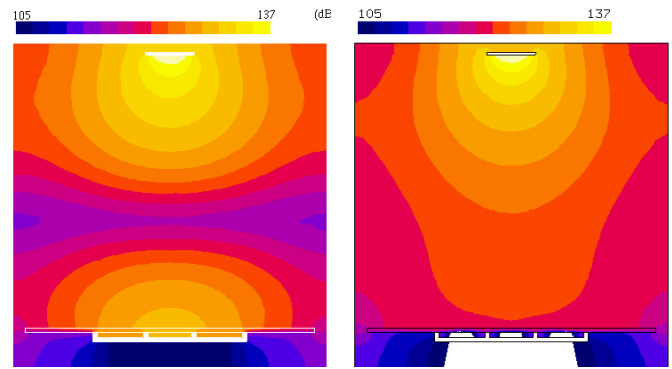

Fig. 8. Acoustic pressure in presence of a porous material at 200 Hz. Left: without control, right: with local impedance active control

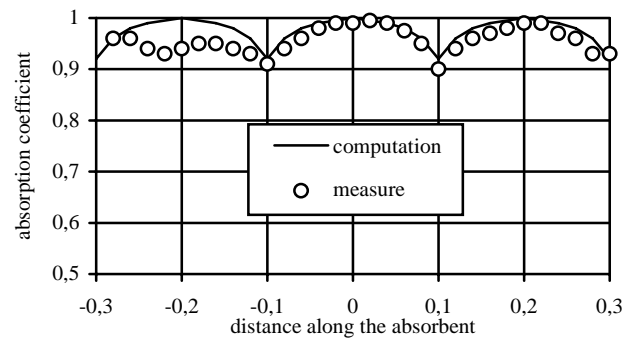

Fig. 9. Absorption coefficients along the absorbent - f = 200 Hz.

REFERENCES

[1] P. Lueg 1936, "Process of silencing sound oscillations," *US patent n° 2043416*.
[2] H.F. Olson & E.G. May 1953, "Electronic sound absorber," *J. Acoust. Soc. Am., 25(6),* pp. 1130- 1136.
[3] P.A. Nelson & S.J. Elliott 1992, *Active control of sound.* London: Academic Press.
[4] M. Furstoss, D. Thenail & M.A. Galland, "Surface impedance control for sound absorption: direct and hybrid passive/active strategies," *Journal of Sound and Vibration, 203(2),* pp. 219- 236, 1997.
[5] L. Nicolas, K.A. Connor, S.J. Salon, B.G. Ruth, L.F. Libelo, "Modélisation 2D par Éléments Finis de phénomènes micro-ondes en milieu ouvert," *J. Phys. III France*, no 2, pp. 2101-2114, 1992.
[6] J.F. Allard, *Propagation of sound in porous media. Modelling sound absorbing materials*, Elsevier Applied Science, 1993.

[7] K.U. Ingard, *Fundamentals of waves and oscillations*, Cambridge University Press, 1988.
[8] J.W.S. Rayleigh, *Theory of Sound*, Dover, New York, 1945.